\documentclass[twocolumn,floatfix,pra,aps,showpacs]{revtex4}
\usepackage{epsfig,graphicx,tabularx}

\begin{document}
%
%\title{Vortices in dipolar Bose-Einstein condensates near Feshbach resonance}
\title{Vortices in Bose-Einstein condensates with dominant dipolar interactions}

\author{M. Abad}
\affiliation{Departament d'Estructura i Constituents de la Mat\`{e}ria,\\
Facultat de F\'{\i}sica, and IN2UB, Universitat de Barcelona, E--08028 Barcelona, Spain}
\author{M. Guilleumas}
\affiliation{Departament d'Estructura i Constituents de la Mat\`{e}ria,\\
Facultat de F\'{\i}sica, and IN2UB, Universitat de Barcelona, E--08028 Barcelona, Spain}
\author{R. Mayol}
\affiliation{Departament d'Estructura i Constituents de la Mat\`{e}ria,\\
Facultat de F\'{\i}sica, and IN2UB, Universitat de Barcelona, E--08028 Barcelona, Spain}
\author{M. Pi}
\affiliation{Departament d'Estructura i Constituents de la Mat\`{e}ria,\\
Facultat de F\'{\i}sica, and IN2UB, Universitat de Barcelona, E--08028 Barcelona, Spain}
\author{ D. M. Jezek   }
\affiliation{Departamento de F\'{\i}sica, Facultad de Ciencias Exactas
y Naturales,
Universidad de Buenos Aires, RA-1428 Buenos Aires, Argentina}
\affiliation{Consejo Nacional de Investigaciones Cient\'{\i}ficas y
T\'ecnicas, Argentina}

\date{\today}
\begin{abstract}
We present full three-dimensional numerical calculations of single
vortex states in rotating dipolar condensates. We consider 
a Bose-Einstein condensate of
$^{52}$Cr atoms with dipole-dipole and
$s$-wave contact interactions confined in an
axially symmetric harmonic trap.
We obtain the vortex states by
numerically solving the Gross-Pitaevskii equation in the rotating frame with
no further approximations. We investigate the properties of a single
vortex and calculate the critical angular velocity
%above which the vortex is energetically favorable
for different values of the $s$-wave scattering length.
%We show that a variational ansatz provides a good description for a
%vortex state for large values of the $s$-wave scattering length.
%However, whereas
%We show that, whereas the variational approach breaks down close 
%to the Feshbach resonance, the exact numerical calculation is a good tool
%to investigate vortex states for different values of the $s$-wave
%scattering length even close to the resonance. 
We show that, whereas the standard variational approach breaks down in the limit of pure dipolar interactions, exact solutions of the Gross-Pitaevskii equation can be obtained for values of the $s$-wave
scattering length down to zero.
The energy barrier for
the nucleation of a vortex is calculated as a
function of the vortex displacement from the rotation axis for
different values of the angular velocity of the rotating trap.

%3D exact numerical results- critical angular frequency- energy
%barriers- core size
\end{abstract}
\maketitle
\section{Introduction}\label{Intro}

%TUNABILITY
%
%the interparticle interaction is dominated by the dipole-dipole forces.
%
%
The experimental realization of a Bose-Einstein condensate of chromium atoms \cite{gri05, Beaufils2008} has encouraged research on the new field of dipolar gases at very low temperature. An ultracold gas of chromium atoms is a very suitable system for
studying dipolar condensates because, in contrast to alkali atoms,
$^{52}$Cr atoms possess an anomalously large magnetic dipole moment.
Moreover, the magnitude and sign of the $s$-wave scattering length
between $^{52}$Cr atoms, and therefore the strength of the contact
interaction, can be experimentally controlled through Feshbach
resonances \cite{Feshbach, Lahaye2007}.% The experimental realization of dipolar Bose-Einstein condensates (BEC)  makes them an appealing system to study.

While the contact interaction is isotropic, the dipolar
potential is anisotropic and long range. The atom-atom interaction
is then determined by the balance of both potentials, giving rise to
interesting phenomena in a dipolar Bose-Einstein condensate (BEC). One of them is their stability, which in contrast to s-wave condensates crucially depends on the trap geometry.
%\cite{Santos2000,Ronen2007,Dutta2007}
In addition, there are other factors that affect the stability of the condensate, such as the scattering length, the magnetic (or electric) moment of the atoms and the number of trapped dipoles.
%\cite{Goral2000}.
%These parameters, together with the magnetic moment of the atoms, control the relative strength of the dipolar interaction to the contact interaction and the trapping potential and they determine the stability conditions of the BEC. 
The problem of stability and collapse in dipolar condensates has
been the subject of intensive experimental \cite{Koch2008, Metz2009, Lahaye2008} and theoretical investigations \cite{Santos2000, Dell2004, Parker2009, Dutta2007, Ronen2007}.% In particular, due to the dependence of the particle-particle interaction on momentum, a roton-maxon excitation spectrum is expected for these systems~\cite{Santos2003, Ronen2007, Dell2003} and their collapse has been predicted to be brought about by the roton mode.

Another interesting feature is the appearance of new structured biconcave ground states
for certain values of the strength of the dipolar interaction and
the harmonic trap anisotropy \cite{Ronen2007,Dutta2007}.
In contrast to condensates with the maximum of the density at the center, which have a roton-like excitation spectrum, these biconcave condensates become unstable due to angular excitations \cite{Ronen2007}.

%It has been pointed out that these biconcave condensates become unstable due to angular excitations, whereas condensates with normal shape, which exhibit the maximum of the density at the center, have a roton-like excitation spectrum.

%Algo de dipolares+ vortices
%
An important issue that is presently under investigation is
the superfluid character of dipolar condensates.
The presence of quantized vortices is a clear signature of
superfluidity in quantum systems \cite{don91}.
Dipolar condensates constitute a unique testing ground of the interplay between different interatomic interactions in the superfluid properties. Since the $s$-wave scattering
length $a$ can be experimentally controlled,
it is appealing to study
vortex states in different regimes, going from a pure dipolar (i.e. $a=0$) to a
pure contact interaction condensate, passing
through BECs with both $s$-wave and dipolar interactions. As $a$ tends to zero, the dipolar interaction becomes comparatively stronger and its effect on the nucleation of vortices is enhanced.

Vortices
in dipolar condensates have been studied within the Gross-Pitaevskii
framework~\cite{wil08,Yi2006,dell07}. It has been shown
that the presence of vortex states affects the stability of a
dipolar condensate \cite{wil08,Yi2006} and that the effect of
dipolar interactions on the critical angular velocity depends on the
geometry of the trap \cite{dell07}. In Ref.~\cite{Yi2006} the
authors have focused on attractive contact interactions in quasi-two-dimensional
rotating dipolar condensates, whereas in Ref.~\cite{wil08} an
axially symmetric non-rotating condensate has been studied. In Ref.~\cite{dell07} a variational approach has been used to describe a
vortex state in the dipolar Thomas-Fermi (TF) limit.

The stability of vortices in dipolar condensates has been recently adressed. In Ref.~\cite{Wilson2009}, the authors have studied the stability and excitations of singly and doubly quantized vortices in dipolar BECs, while in Ref.~\cite{Klawunn2009} a phase transition has been predicted between straight and twisted vortex lines. Also, the transverse instability of vortex lines has been studied in Ref.~\cite{Klawunn2008}.

In this work we consider vortex states in
three-dimensional (3D) rotating dipolar
condensates. Our aim is to investigate the
effect of the dipolar interaction on the vortex properties when the contact interaction is low enough to consider that the dipole-dipole interaction is dominant. To this end, we use a full 3D approach to numerically solve the Gross-Pitaevskii equation (GP). We concentrate on dipolar condensates with $s$-wave scattering length approaching zero and,
%We center on condensates with $s$-wave scattering length approaching zero and a pancake geometry %
in particular, we study the structure of the vortex core and obtain the critical rotation
frequency necessary to nucleate a vortex.
We also evaluate the energy barrier for vortex formation
as a function of the vortex distance from the trap center.
%We characterize the core size.

This work is organized as follows. In Sec.~\ref{Theory} we describe the
theoretical framework and the system under study.
In Sec.~\ref{GS} we revisit some properties of the ground state of dipolar BECs.
In Sec.~\ref{VS} we investigate the formation of a centered vortex
in a rotating frame, and in Sec.~\ref{Barrier} we present the calculation of the nucleation barrier.
Finally, a summary and concluding remarks are
offered in Sec.~\ref{Conclusions}.

\section{Theoretical framework}\label{Theory}
%\section{The system}

We consider a Bose-Einstein condensate of $N$ chromium atoms
at zero temperature,
confined by an axially symmetric harmonic potential
\begin{equation}
V_{\text{trap}}(\mathbf{r})=
  \frac{m}{2}\, (\omega_{\perp}^2 r_\perp^2 +\omega_{z}^2 z^2) \,,
\end{equation}
where $r_\perp^2=x^2+y^2$, $m$ is the atomic mass, and $\omega_{\perp}$
and $\omega_{z}$ are the radial and axial angular trap frequencies,
respectively.
The aspect ratio of the trapping potential
is $\lambda=\omega_z/\omega_\perp$. We shall only consider pancake shape condensates, in which case $\lambda >1$ \cite{foot1}.

Since $^{52}$Cr has a large magnetic dipole moment, $\mu
= 6 \mu_B$ ($\mu_B$ is the Bohr magneton), chromium atoms interact
not only via $s$-wave contact interactions but also via the
dipole-dipole interaction, which can be written as:
\begin{equation}
v_{\text{dip}} (\mathbf{r}-\mathbf{r'})= \frac{\mu_0 \mu^2}{4 \pi}
\frac{1 - 3 \cos^2 \theta}{|\mathbf{r}-\mathbf{r'}|^3} \,,
 \label{dip-pot}
\end{equation}
where $\mu_0$ is the vacuum permeability,
$\mathbf{r}-\mathbf{r'}$ is the distance between the dipoles, and
$\theta$ is the angle between
the vector $\mathbf{r}-\mathbf{r'}$ and the dipole axis, which we take to be $z$. In this configuration the dipoles are situated head to tail along the $z$ axis. % and side by side in the $xy$ plane.
The dipolar interaction is then attractive along the magnetization direction and repulsive in the perpendicular one. Since we are considering a pancake geometry, this interaction is mainly repulsive.

In the mean-field framework, the GP equation
provides a good description of a weakly interacting dipolar BEC,
provided that the dipolar interaction is not too large.
%
%\cite{Menotti07}.
%
In order to investigate vortex states, we assume that the
condensate is rotating around the symmetry axis $z$ with angular
frequency $\Omega$. Using the imaginary-time propagation method in 3D we obtain the solutions of the GP equation in the rotating frame:
\begin{eqnarray}
 & &
 \left[ -\frac{ \hbar^2}{2m} \nabla^2 + V_{\text{trap}} +
g \, |\psi(\mathbf{r})|^2 \right. + \left. V_{\text{dip}}(\mathbf{r})
-\Omega \hat{L}_z\right] \psi(\mathbf{r})= \nonumber \\
%&&+ \left. \frac{\mu_0 \mu^2}{4 \pi} \int d\mathbf{r'} \,
%\frac{1 - 3 \cos^2 \theta}{|\mathbf{r}-\mathbf{r'}|^3} \,|\psi(\mathbf{r'})|^2
&& 
 \qquad\qquad= \tilde{\mu} \, \psi(\mathbf{r}) \,,
\label{gp}
\end{eqnarray}
where $\psi(\mathbf{r})$ is the
condensate wave function normalized to the total number of
particles, $\hat{L}_z$ is the angular momentum operator along the $z$ axis and $\tilde{\mu}$ is the chemical potential. The contact interaction potential is characterized by the
coupling constant $g=4\pi\hbar^2 a /m$ and the mean-field dipolar interaction $V_{\text{dip}} (\mathbf{r})$ is given by:
\begin{equation}
 V_{\text{dip}} (\mathbf{r})= \int d\mathbf{r'} v_{\text{dip}} (\mathbf{r}-\mathbf{r'}) |\psi(\mathbf{r'})|^2\ . \label{Vdip}
\end{equation}
%
%A singly quantized vortex line along the $z$ axis can be described
%by writing the wave function in the form $\psi({\bf r})=\psi_v({\bf
%r}) \exp(i \phi)$, where $\phi$ is the azimuthal angle around the
%$z$ axis. This state carries an angular momentum along $z$ which is
%$L_z= N \hbar$.
%By imposing this ansatz on the wave function, the Gross-Pitaevskii
%equation for a vortex state takes the form:
%\begin{eqnarray}
% & &\mu \, \psi_v(\mathbf{r}) =
% \left[ -\frac{ \hbar^2}{2m} \nabla^2 + \frac{\hbar^2}{2 m r_\perp^2}
% +V_{\text{trap}} +
%g \, |\psi_v(\mathbf{r})|^2 \right. \nonumber \\
%&&+ \left. \frac{\mu_0 \mu^2}{4 \pi} \int d\mathbf{r'} \,
%\frac{1 - 3 \cos^2 \theta}{|\mathbf{r}-\mathbf{r'}|^3} \,
%|\psi_v(\mathbf{r'})|^2 \right] \psi_v(\mathbf{r}) \,,
%\label{gp}
%\end{eqnarray}
%where $\mu$ is the chemical potential.
%The contact interaction potential is characterized by the
%coupling constant $g=4\pi\hbar^2 a /m$, where $a$ is the $s$-wave
%scattering length. The condensate wave function is normalized to the
%total number of particles.
%If a vortex is present in the condensate, the atoms flow around
%the vortex core with quantized circulation, which yields the
%centrifugal term in Eq.~(\ref{gp}).
%
%
%\begin{equation}
% U_{\text{dip}} (\mathbf{r},\mathbf{r'})= \frac{\mu_0 \mu^2}{4 \pi}
%\frac{1 - 3 \cos^2 \theta}{|\mathbf{r}-\mathbf{r'}|^3} \,,
% \label{dip-pot}
%\end{equation}
%where $\mu_0$ is the vacuum permeability, and $\theta$ is the angle
%between
%%the dipole moment and the vector $(\mathbf{r}-\mathbf{r'})$.
%the vector $(\mathbf{r}-\mathbf{r'})$ and the dipole axis.
%
When the dipolar BEC is at rest, the ground state is obtained from
Eq.~(\ref{gp}) by setting $\Omega=0$.

% ******* ENERGY FUNCTIONAL *********
%The energy density functional has the standard form but
%with a new term, $E_{\text{dip}}$, which is the interaction energy
%due to dipole-dipole interactions:
%\begin{equation}
%  E [\psi] \! =\! \int \! \left( \frac{ \hbar^2 }{2 m}  |\nabla \psi |^2 +
%V_{\text{trap}} \,|\psi|^2 + \frac{g}{2} \, |\psi|^4 \right)
% d {\bf r} + E_{\text{dip}} \,, \label{ed}
%\end{equation}
%where
%\begin{equation}
%E_{\text{dip}}=\frac{1}{2} \int V_{\text{dip}}(\mathbf{r}) \,
%|\psi(\mathbf{r})|^2 d\,\mathbf{r}\,.
%\end{equation}

The dipolar term in Eq.~(\ref{gp}) transforms the usual GP
equation in a more complicated integro-differential equation.
However, the dipolar
interaction integral (\ref{Vdip}) can be evaluated by using fast-Fourier transform (FFT) techniques \cite{Goral2002} and the introduction of a cutoff at small distances \cite{Goral2000}. We have used the FFTW package \cite{FFTW} to compute the discrete Fourier transforms.

The energy density functional in the rotating frame has the standard GP form but
with a new term, $E_{\text{dip}}$, which is the interaction energy
due to the dipole-dipole potential:
\begin{eqnarray}
 E [\psi]  && = E_{\text{kin}} + E_{\text{trap}} + E_{\text{int}} + E_{\text{dip}} + E_L= \nonumber\\
 && \hspace{-0.8cm}= \int\frac{ \hbar^2 }{2 m}  |\nabla \psi |^2 d {\bf r} + \int V_{\text{trap}} \,|\psi|^2 d {\bf r} + \int\frac{g}{2} \, |\psi|^4 d {\bf r} +   \nonumber\\
&& \hspace{-0.8cm} + \frac{1}{2} \int V_{\text{dip}} \,
|\psi|^2 d\mathbf{r}\, - \Omega\int\psi^* \hat{L}_z\psi\, d\mathbf{r}.
\label{ed}
\end{eqnarray}
%The virial theorem can be obtained from this expression and constitutes a good check of the accuracy of the numerical solution. It has already been shown by using the principle of scale invariance \cite{Jezek2004} that defining the transformation $\textbf{r}\rightarrow\nu\,\textbf{r}$ the three first contributions to the energy (\ref{ed}) scale following the rules: $E_{\text{kin},\nu} = \nu^2 E_{\text{kin}}$, $E_{\text{trap},\nu}=E_{\text{trap}}/\nu^2$ and $E_{\text{int},\nu}=\nu^3E_{\text{int}}$. 
%The condition that the scaling transformation must preserve the norm of $\psi$ leads to the scaling of the condensate wave function: $\psi_\nu({\bf r})=\nu^{3/2}\psi({\bf r})$. One can calculate then the contribution of the dipolar energy:
%\begin{eqnarray}E_{\text{dip},\nu} &&\!\!\!= \frac{1}{2}\frac{\mu_0\mu^2}{4\pi}\!\!\int\! d{\bf r} d{\bf r'} \frac{\nu^6|\psi(\nu{\bf r})|^2|\psi(\nu{\bf r'})|^2}{|{\bf r}-{\bf r'}|^3}(1-3\cos^2\theta)\nonumber\\
%\!\!\!\!=\!\frac{1}{2}&&\!\!\!\!\!\!\!\frac{\mu_0\mu^2}{4\pi}\nu^3\!\!\int\! d(\nu{\bf r}) d(\nu{\bf r'}) \frac{|\psi(\nu{\bf r})|^2|\psi(\nu{\bf r'})|^2}{|\nu{\bf r}-\nu{\bf r'}|^3}(1-3\cos^2\theta)\nonumber\\
%&&\!\!\!\!\!\!\!\!\!\!\!\!\!\!\!\!\!=\nu^3 E_{\text{dip}}\ .
%\end{eqnarray}
%
%Therefore the total energy of the system can be written as:
%\[E_\nu= \nu^2 E_{\text{kin}} + \frac{1}{\nu^2}E_{\text{trap}} + \nu^3E_{\text{int}} + \nu^3 E_{\text{dip}}\ .\]
%Imposing that this expression is minimum for $\nu=1$, 
From scaling considerations (see Appendix \ref{Virial}), one can show that the virial theorem for a dipolar condensate reads:
\begin{equation}
 2E_{\text{kin}} - 2E_{\text{trap}} + 3E_{\text{int}} + 3E_{\text{dip}} = 0 \ .\label{EqVirial}
\end{equation}
The fulfillment of this expression constitutes a good check of the accuracy of the numerical solution. In all the numerical calculations presented here we have checked that the condition (\ref{EqVirial}) is well satisfied.% with a maximum deviation from zero of the order of $3 \%$ relative to the total energy.

For purely dipolar condensates,
it is useful to introduce a dimensionless parameter
%to characterize dipolar interaction strength \cite{Wilson08}
that measures the effective strength of the dipolar interaction \cite{Ronen2007}:
\begin{equation}
 D =  \frac{ N  \, \mu_0 \, \mu^2 \, m}{4 \pi \, \hbar^2 \, a_{\perp}} \,,
\label{d}
\end{equation}
where $a_{\perp} = \sqrt{ \hbar / (m \omega_\perp)}$ is the transverse
harmonic oscillator length that characterizes the radial mean size of the
noninteracting condensate.
Increasing the value of $D$ is equivalent to
increase the number of atoms in the trap or their dipolar moment. %When a critical value $D_{c}$ depending on the trap aspect ratio is reached the condensate becomes unstable \cite{Ronen2007}.

For dipolar condensates with also contact interactions, one can introduce another dimensionless quantity
%parameter
to characterize the relative strength of the
dipolar and $s$-wave interactions \cite{Martikainen2001}:
\begin{equation}
\epsilon_{dd}=\frac{\mu_0 \mu^2 m}{12 \pi \hbar^2 a} \,.
\label{epsilondd}
\end{equation}
This parameter is defined in such a way that a homogeneous BEC with
$\epsilon_{dd} > 1$ is unstable \cite{Menotti2007}. It has been
shown~\cite{Eberlein2005} that in the Thomas-Fermi limit 
%(when $N a/a_\perp \gg 1$)
a dipolar BEC also becomes unstable when $\epsilon_{dd} > 1$. In this work we mainly study systems with $a\rightarrow0$ that are far from the TF limit. This means that $\epsilon_{dd}$ does not need to be smaller than one for the BEC to be stable and indeed we still find stable solutions for $\epsilon_{dd}>1$.

\section{Ground state}\label{GS}
Before studying vortex states it is interesting to characterize
the ground state wave function of the system,
$\psi_0(\mathbf{r})$. It can be obtained from the minimization of the total energy
(\ref{ed}), that is, by solving the GP equation (\ref{gp}) in the
laboratory frame ($\Omega=0$).  As a starting point, we have checked
that the results of our full 3D calculation are in agreement with
the stability diagram of a pure dipolar condensate in a pancake trap previously
calculated \cite{Ronen2007}.%; we have obtained
%condensates with a biconcave shape for the same parameters as in Ref.~\cite{Ronen2007}.

%As it was said in the introduction, 
%the dipole-dipole interaction is anisotropic.

The main effect of the dipolar interaction on the ground state of
the system is to deform the condensate as compared to the $s$-wave
case. This effect is a direct consequence of the anisotropy of the
dipolar potential and depends on the specific trap that is
considered. For a spherical trap, the effect of the dipolar
interaction is to squeeze the cloud along the repulsive direction while stretching it in the
attractive direction. Although this might be somewhat counter-intuitive, it
is easily explained taking into consideration the particular shape
of the dipolar potential (see, for instance, Ref.~\cite{Stuhler}):
since the dipolar potential shows a saddle configuration with two minima along the magnetization
axis (attractive direction) it is less expensive for the system to
accommodate more dipoles along this direction than along the repulsive one,
thus the cloud size becomes larger in the former and
smaller in the latter.

\begin{figure}
\epsfig{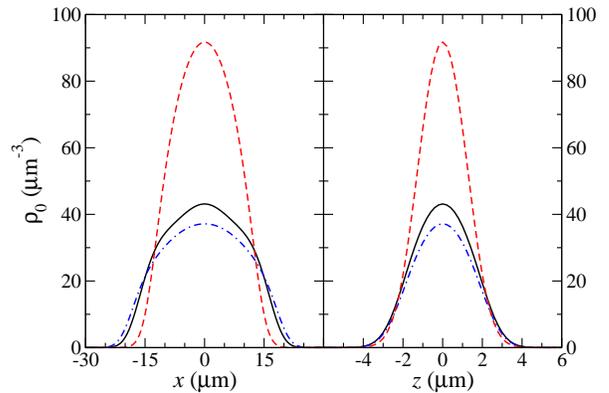} \caption{(Color online) Ground
state densities of pure $s$-wave (dashed line, with $a=5\,a_B$), pure
dipolar ($\mu=6\mu_B$, solid line) and both $s$-wave and dipolar
($a=5\,a_B$, $\mu=6\mu_B$, dot-dashed line) condensates. They all
correspond to the case $D=50$, $\lambda=11$, with $N=10^5$ and
$\omega_\perp=8.4\times2\pi$ s$^{-1}$.} \label{density0}
\end{figure}

However, the deformation of the condensate is different for
non-spherical traps. In Fig.~\ref{density0} we show the density
profiles in the perpendicular and
parallel directions to the magnetization axis ($x$ and $z$, respectively) for three different
condensates: a pure dipolar BEC, a pure $s$-wave BEC ($a=5\,a_B$)
and one with the two types of interactions. All of them contain $N=10^5$ bosons in a trap with
frequencies $\omega_\perp=8.4\times2\pi$ s$^{-1}$ and
$\omega_z=92.5\times2\pi$ s$^{-1}$.
The asymmetry parameter of such a configuration is $\lambda=11$ and the dipolar
parameters take the values $D=50$ and $\epsilon_{dd}=3.03$.

In order to quantitatively analyze the deformation of the condensate, we compute the root-mean-square radii in the radial and axial directions, namely $R_\perp$ and $R_z$. Figure~\ref{density0} shows that the deformation is different to the spherical case: the gas is stretched in both
directions of space, even though the cloud aspect ratio
$\kappa=R_\perp/R_z$ increases ($\kappa=8.28$ for the $s$-wave case,
$\kappa=9.06$ for the purely dipolar case and $\kappa=9.65$ when
both interactions are considered) \cite{foot2}. 
The difference now is that the condensate is
tightly trapped in the $z$ direction so that the energy decrease achieved by putting more and more dipoles along the
magnetization direction has to overcome a stronger trapping
potential. The result of this energetic balance is that the system cannot accommodate so many
atoms along the $z$ axis, hence the cloud has to stretch in all directions.

%Following the same line of reasoning, a dipolar condensate in a cigar-shaped trap with the magnetization axis parallel to the trap axis would be even more elongated than a $s$-wave condensate in the same geometry. However, in this case we have to deal with a further impediment: since in such a trap the dipolar potential is mainly attractive, the restriction on the number of particles that can form a stable BEC is much stronger.

\begin{figure}
\epsfig{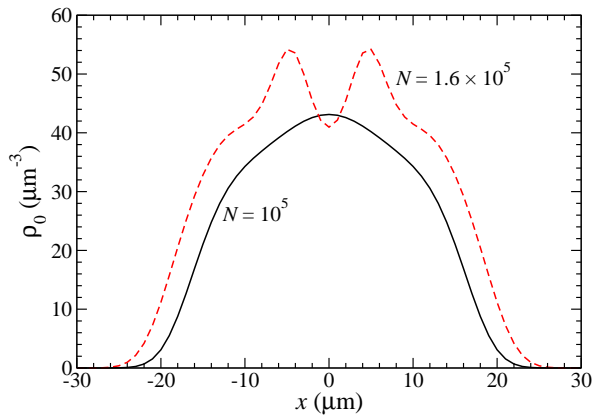}
\caption{(Color online) Density profile for the
 ground state configuration of pure dipolar BECs cointaining $N=10^5$ (solid line) and
$N=1.6 \times 10^5 $  (dashed line)
 atoms in the trap with $\omega_{\perp}=8.4 \times 2 \pi$ s$^{-1}$ and
$\omega_z= 92.5 \times 2 \pi$ s$^{-1}$.
 }
\label{density}
\end{figure}

On the other hand, the ground state of a dipolar condensate may present some stable density
structures with the density maximum away from the center. They are usually found in isolated regions of the
parameter space ($D,\lambda$)
that are close to instability. This means that by increasing a little the value of $D$ the condensate enters the unstable regime and overcomes a collapse which is thought to be of angular type \cite{Ronen2007}. Figure~\ref{density}
shows the density profiles of two different stable ground state configurations of a pure
dipolar ($a=0$) condensate, one of them having a normal shape, while the other shows
 a biconcave structure. They both correspond to a condensate confined in a harmonic
 potential with asymmetry $\lambda=11$ ($\omega_\perp = 8.4 \times
2 \pi$ s$^{-1}$ and $\omega_z = 92.5 \times 2 \pi$ s$^{-1}$), but
two different numbers of trapped atoms $N=10^5$ (solid line) and
$1.6 \times 10^5$ (dashed line), which correspond to dipolar
interaction parameters $D=50$ and $80$, respectively.

%As can be appreciated in the density profiles in Fig.~\ref{density}, the solid line
%($D=50$) corresponds to a condensate with a normal shape with the
%maximum of the density in the center of the trap, whereas the dashed
%line ($D=80$) corresponds to a biconcave configuration with a local
%minimum of the density in the center of the trap and its maximum
%displaced from the center. 

It is important to note that
although the parameter $D$ is well suited to determine the
regions where the density presents a biconcave shape,
%density-structured regions,
 it fails to characterize all the physics
underlying purely dipolar condensates. 
Clearly, we need at least three
parameters to properly describe such systems since we have three
degrees of freedom, namely $N$, $\omega_\perp$ and $\omega_z$.

%As it can be appreciated in the column density, the solid line ($D=50$)
%corresponds to a condensate with a normal shape with the maximum of the density
%in the center of the trap, whereas the dashed line ($D=80$) corresponds to a
%biconcave configuration with a local minimum of the density in the center of the
%trap and its maximum displaced from the center.

%Fig.~\ref{density}  shows the column density for different ground state
%configurations of a pure dipolar condensate ($a=0$). We plot the particle
%density integrated along the $x$ direction, $n_0(x)=\int dz \,|\psi_0(x,0,z)|^2$,
%instead of the density profile because it is a typical measured quantity. We
%have considered a pancake-shaped trap with $\lambda=11$
%($\omega_\perp = 8.408 \times 2 \pi$ Hz and $\omega_z = 92.491 \times 2 \pi$ Hz),
%and two different number of trapped atoms $N=100000$ (solid line) and $160000$
%(dashed line), that correspond to a dipolar interaction parameter $D=60$ and $80$,
%respectively. As it can be appreciated in the column density, the solid line ($D=60$)
%corresponds to a condensate with a normal shape with the maximum of the density in the
%center of the trap, whereas the dashed line ($D=80$) corresponds to a biconcave
%configuration with a local minimum of the density in the center of the trap and
%its maximum displaced from the center.

%We consider a pancake-shaped trap ($\lambda > 1$) with the dipoles polarized along the $z$ axis. In this configuration the dipoles are situated mainly side-by-side. Therefore, the dipolar interaction is predominantly repulsive and the dipolar BEC is stable. However, when the number of dipoles is large enough, and the dipolar interaction parameter $D$ exceeds a certain threshold depending on the trap aspect ratio, the dipolar condensate becomes unstable.  As a starting point, we have checked that the results of our full 3D calculation are in agreement with the stability diagram of a pure dipolar condensate in a pancake trap calculated previously in 2D \cite{Ronen2007}. We have also obtained condensates with a biconcave shape in the same isolated regions of the parameter space found in Ref.~\cite{Ronen2007}. These shapes are a pure dipolar effect, mainly due to the anisotropic character of the interaction. The physical mechanism that triggers the instability in this isolated regions is still under discussion \cite{Ronen2007}.

%*** Ver si vale la pena incluir un grafico asi. O junto con un vortice. ***

\section{Vortex states}\label{VS}
\subsection{Critical rotation frequency and vortex generation}
%
%\section{ Vortex generation and critical rotation frequency}

%Muntsa: no me acuerdo porque no pusieron  un termino centrifugo
%directamente para generar el vortice centrado.
%Aca el constraint cuadratico para
%generar el vortice no me parece que convenga introducirlo.
%Vos que decis, a mi me parece
%que podemos meternos en lios, es decir que el referee puede
%objetar la forma de generarlos.
%Si no hay una razon fuerte podriamos decir que lo generarmos con
%un termino centrifugo,
%o directamente no decir nada. y continuar como sigue:

%In a non-rotating condensate
The inclusion of vorticity is accompanied by an energy cost due to
the appearance of angular momentum.
%Thus in order to generate vortices one may minimize the energy
%(\ref{ed}) in a rotating frame \cite{dal99}
%
Thus, in order to generate vortices the condensate must rotate. In a
frame rotating at an angular frequency $\Omega$ about the $z$ axis, the
energy of the condensate carrying angular momentum $L_z$ becomes
$(E-\Omega L_z)$, where $E$ and $L_z$ are evaluated in the
laboratory frame. At low rotation frequencies this energy is
minimal without the vortex (ground state configuration). But if
$\Omega$ is large enough the creation of a vortex can become
favorable due to the $-\Omega\hat{L_z}$ term. This happens at the
critical frequency $\Omega_c$.
The thermodynamical critical angular velocity for nucleating a
singly quantized vortex is obtained by subtracting from the
vortex state energy $E_v$ in the laboratory frame the ground-state
energy $E_0$, i.e.,
$\Omega_c =  ( E_v - E_0)/ N \hbar$, and it provides a lower bound
to the experimental critical angular velocity~\cite{Dalfovo1999, Fetter2001}.
%
%\begin{equation}
% H ( \Omega ) = H - \Omega L_z.
%\label{gp}
%\end{equation}
%

Depending on the value of $\Omega$, condensates with different
number of vortices can be obtained, %
from a single vortex
configuration at low angular velocities ($\Omega \sim \Omega_c$) to
vortex lattices at high rotation frequencies. When the rotation
frequency is close to the radial frequency of the trap the number of
vortices becomes so large that the distance between them is
smaller than the vortex cores, entering the strongly interacting
regime~\cite{LLL}.
%
%Note that there exists an upper bound for the angular velocity of
%rotation of the trap given by the radial frequency
%$\omega_\perp$ of the confining potential. For higher values the
%system is no longer confined.
%
In this work we are interested in single vortex
configurations that are favorable when the angular velocity slightly
exceeds the critical value.

%Numerical results: imprinting phase and with no initial ansatz for
%the vortex wave function... same results.
%
%For a given value of the scattering length and fixed an angular
%frequency $\Omega$, we have solved the GP equation in the
%corresponding rotating frame (\ref{gp}) to obtain the configuration
%that minimizes the energy. We have used different initial wave
%functions to start the numerical iteration process to avoid any bias
%to the final solution. Both, an initial vortex free configuration
%and an off-centered vortex line imprinted by using the ansatz
%(\ref{psi_v}) lead to the same converged solution. We have checked
%the quantization of the circulation around the vortex core of our
%vortex solutions.
%
%
%As expected, for small values of $\Omega$ the ground state
%corresponds to a free-vortex configuration, but for values of the
%angular frequency equal or slightly larger than a critical one, the
%centered vortex state is the one that minimizes the energy. From
%these results we have obtained the critical angular frequency
%$\Omega_c$ for different values of the scattering length.

We have numerically
computed a vortex state by imprinting a phase
to the initial wave function and solving
the GP equation (\ref{gp}) in the rotating frame
%This method consists in minimizing the energy,
%at a given value of the angular velocity $\Omega$, starting
%from a convenient guess for the initial wave function, but
without fixing the vorticity during the minimization process.
We have used the following ansatz for the initial wave function~\cite{Jezek08}:
\begin{equation}
\psi({\bf r})=\psi_0({\bf r}) \, \frac{x+iy}{\sqrt{x^2+y^2}} \,,
%\psi({\bf r})=\psi_0({\bf r}) \, e^{i \phi} \,,
%\psi({\bf r})=\psi_0({\bf r}) \exp(i \phi)\,,
\label{wf-ansatz}
\end{equation}
where $\psi_0$ is the ground state wave function.
From this ansatz, it is straightforward to see that $\psi$ and
$\psi_0$ have the same density profile, but $\psi$ has an imprinted
velocity field which is irrotational everywhere except at the
vorticity line ($z$ axis). %Therefore, the initial wave function
%$\psi$ is obtained by phase imprinting to the original nonvortex
%state $\psi_0$. 
Moreover, $\psi$ is an eigenstate of $\hat{L_z}$ with eigenvalue $N \hbar$.
%about $z$.
%

Solving the GP equation~(\ref{gp}), we have obtained, as expected, that for
small values of $\Omega$ the system converges to a vortex-free
configuration that corresponds to the ground state.
However, for
values of the angular frequency equal or slightly larger than the
critical one, a centered vortex state configuration minimizes the
energy.
%From these results we have obtained the critical angular frequency
%$\Omega_c$ for different values of the scattering length.
We have checked that the circulation is quantized around the
vorticity line.

\begin{figure}
\epsfig{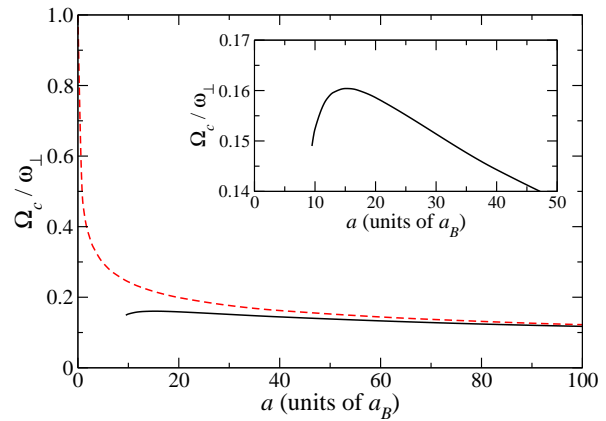}
\caption{(Color online) Critical angular velocity above which a singly quantized
vortex is energetically favorable in a pancake trap with aspect
ratio $\lambda=5$, as a function of the $s$-wave scattering length.
The solid line corresponds to $s$-wave plus dipolar interactions.
The dashed line corresponds to only $s$-wave contact interaction. Inset: behavior of $\Omega_c(a)$ corresponding to both
$s$-wave and dipolar interactions
%around the resonance
close to the instability limit.
 }
 \label{omega-5}
\end{figure}
The value of the critical angular frequency $\Omega_c$ above which a
vortex state is energetically favorable depends on the interaction
parameters (scattering length, dipole moment), as well as on the
number of atoms and on the trap geometry.
We plot in Fig.~\ref{omega-5} the critical angular velocity for
vortex nucleation
%$ \Omega_c / \omega_\perp $
as a function of the scattering length, for a condensate with $
N=1.5 \times 10^5 $,
$\omega_\perp=2 \pi \times 200$ s$^{-1}$, $ \lambda = 5$ and
$D=365.68$. 
These are the same parameters as in
Ref.~\cite{dell07}, where a variational ansatz was used to describe the vortex solution. By directly solving the dipolar GP equation without any further approximation we can go below the limit $a=17.5\,a_B$ imposed by their variational ansatz and reach smaller scattering lengths.
%For values of the scattering length below $a=9.5 \,a_B$, the dipolar interaction becomes too repulsive and the system collapses. 
However, for $a<9.5 \,a_B$ the $s$-wave repulsion in the $z$ direction is not strong enough to balance the attraction brought about by the dipole-dipole interaction and the
system becomes unstable.

%Here we use the same parameters as in
%Ref.~\cite{dell07} and go beyond the limit imposed by their variational ansatz.
%
%to compare their variational ansatz for the vortex state wave
%function with our 3D calculation.
%

As a reference we have also calculated the critical angular velocity
necessary to nucleate a vortex in a non-dipolar condensate (with
only $s$-wave interaction), see dashed curve in Fig.~\ref{omega-5}.
In the noninteracting limit ($a=0$) a vortex state corresponds to the first
excited state with energy $E_v=E_0+ N \hbar \omega_\perp$, and
therefore the critical angular frequency is $\Omega_c=\omega_\perp$
\cite{Fetter2001}. The inclusion of the
dipolar interaction (solid curve) causes a decrease of the
critical angular frequency,
which becomes more sizeable for small $s$-wave interactions. % (see also 
%Fig.~\ref{omega-11}).
%
As has been already pointed out \cite{dell07}, it follows from
Fig.~\ref{omega-5} that in a pancake trap
%effect of dipolar interactions in a pancake trap is to decrease the
%value of $\Omega_c$, that is,
it is easier to nucleate a vortex in the presence of dipolar
interactions. 
This can be understood as follows: 
in a pancake shaped condensate, the interaction
between dipoles aligned along the $z$ axis is repulsive on average, hence the maximum
density diminishes with
respect to the pure $s$-wave value (see the variation of the central
density in Fig.~\ref{density0}) and it becomes easier to take the
atoms away from the $z$ axis to nucleate a centered vortex. 
Thus,
the effect of dipolar interactions is to decrease the critical
angular frequency of vortex nucleation in a pancake shaped
condensate.
%
%Conversely, in a cigar shaped condensate with the dipoles aligned
%along the symmetry axis, the dipole-dipole interaction is on average
%attractive. And therefore, the dipolar interactions
%increase the value of $\Omega_c$ \cite{dell07}.
%

%For the set of parameters of Fig.~\ref{omega-5} the dipolar
%condensate becomes unstable for small values of the $s$-wave
%scattering length. The minimum value used in Ref.~\cite{dell07} is
%$a=17.5 a_B$ because the variational approximation breaks down for
%smaller values. Instead, in the present 3D calculation one may go down to $ a= 9.5 a_B$. 
We have found a good agreement between our results and those in Ref.~\cite{dell07} for large values of the scattering length, while for
small values our $\Omega_c(a)$ curve exhibits a
maximum around $a= 15 a_B$ (see the inset in Fig.~\ref{omega-5}). 
The maximum value of the critical angular frequency 
%that the system attains in the condensate of Figure \ref{omega-5} 
is about $\Omega_c^{max}\simeq 0.16~ \omega_\perp$ at $a=15 a_B$, and
corresponds to a dipolar condensate with $\epsilon_{dd}=1.01$
and $D=365.68$.
Decreasing further the scattering length %, before the instability occurs, 
the critical angular velocity necessary to nucleate a vortex
also decreases, in contrast to the general trend at large $s$-wave
values.

\begin{figure}
\epsfig{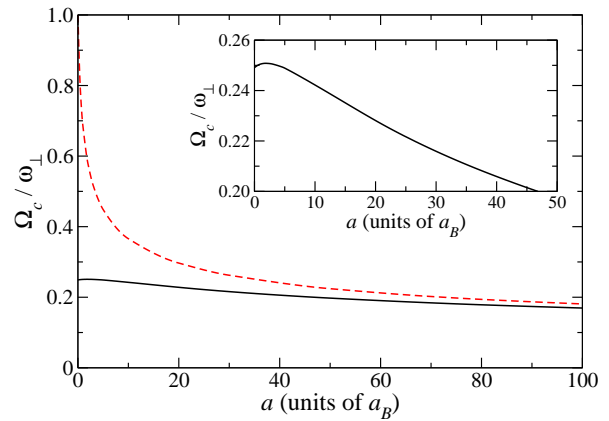}
\caption{(Color online) Critical angular velocity for a singly quantized
vortex in a pancake trap with aspect
ratio $\lambda=11$, as a function of the $s$-wave scattering length.
The solid line corresponds to $s$-wave plus dipolar interactions.
The dashed line corresponds to only $s$-wave contact interaction. Inset:
behavior of $\Omega_c(a)$ around $a\sim 0$ for a
condensate with $s$-wave plus dipolar interactions.
\label{omega-D}
 }
\end{figure}

The presence of a maximum value in the $\Omega_c(a)$ curve is a
consequence of the balance between contact and dipolar interactions.
To explore the origin of $\Omega_c^{max}$ we have
also studied the critical angular velocity as a function of the
$s$-wave scattering length for a system that remains stable at
vanishing values of the scattering length, which corresponds to a pure dipolar
condensate. We plot $\Omega_c$ in Fig.~\ref{omega-D} for $ D= 50
$, $ \omega_\perp= 2 \pi \times 8.4 $ s$^{-1}$, $ \lambda= 11 $, and $ N =
10^5 $. A detail of the behavior of $\Omega_c(a)$ corresponding to
contact plus dipolar interactions for small values of the
$s$-wave scattering length is shown in the inset.
A maximum value of $\Omega_c$ appears also in Fig.\ref{omega-D}
around a scattering length $a \simeq 2 a_B$, which corresponds to a
dimensionless parameter $\epsilon_{dd}=7.57$.
In this case, the effect of the dipolar interaction
%, already mentionedin Fig.~\ref{omega-5},
 becomes
more clear: for a fixed value of the $s$-wave scattering
length, the inclusion of the dipole-dipole interaction decreases the
value of the critical angular velocity for vortex formation. In the
limit of $a=0$, that is for a pure dipolar condensate, $\Omega_c$ is
around a factor $0.25$ smaller than the non-interacting value.

%Note that the value at $ a_s = 0 $ is about a quarter of the noninteracting value.
% Muntsa:  en el caso no interactuante el omega critico da
%directamente el omegar, mi duda es: cuanto mas interactuante mas
%grande es el intervalo entre el omega critico y el omegar ? por lo
%menos es lo que pasa en el regimen de TF, no?
%

\begin{figure}
\epsfig{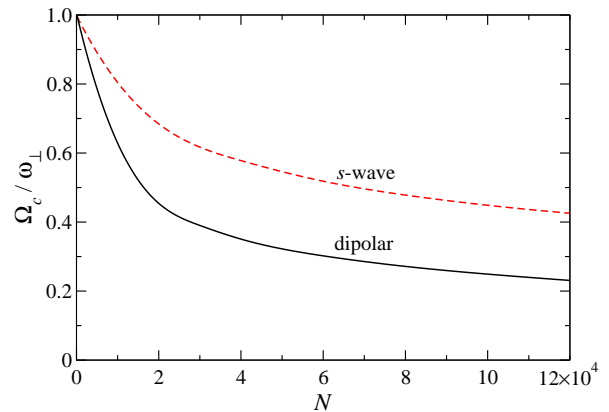}
\caption{(Color online) Critical angular velocity as a function
%of the dimensionless parameter $D$,
of the number of atoms trapped in the same trap as in
Fig.~\ref{omega-D}. The solid line corresponds to a pure dipolar
condensate, and the dashed line to a condensate with only contact
interaction ($a=5 a_B$).} \label{omega-11}
\end{figure}

The critical angular velocity for producing
a vortex is plotted in Fig.~\ref{omega-11} as a function of the
%dimensionless parameter $D$,  Eq.~(\ref{d}),
%Increasing the $D$ value corresponds to increase the number of trapped
%atoms
number of atoms confined in the same trap as in
Fig.~\ref{omega-D}, for a pure dipolar condensate (solid line) and
for a condensate with only contact interaction with $a=5 a_B$
(dashed line). For a given number of atoms the effective repulsion
of the dipolar interaction in a pancake shaped condensate is larger
than the repulsive contact interaction; therefore, $\Omega_c$ is
smaller in a pure dipolar condensate. However, both curves have
a similar behavior: $\Omega_c$ decreases with increasing $N$,
as the repulsion also increases.

%In Fig.~2 we display equidensity contours for a system
%with $N =  \times 10^{5}$ atoms hosting a vortex.
%While in Fig.~3  we show the density profile $\rho_v$ as a
%function of $x$ at $y=z=0$.
%For comparison we show also the corresponding ground state density
%$\rho_0(x,y=0,z=0)$ as a dashed line. In order to stress the main
%differences between the density of the a vortex state and that of the
%groundstate, we depict in the inset of Figs. 3  the
%form factor $ f = \rho_v / \rho_0 $  around a  vortex line.
%Note that $f$ is almost unity outside the vortex core,
%thus the density with and without vortices
%are almost the same except in a narrow region around
%the cores of the vortices.

\subsection{Vortex core size }

The vortex states we have obtained as a result of the 3D
minimization process are straight vortex lines. For a condensate with only contact interaction a characteristic
length for describing the core size of a vortex is given by the
local healing length. In particular, the balance between the kinetic
energy and the interaction energy fixes a typical distance over
which the condensate wave function can heal. For a dilute Bose gas
the healing length is given by $\xi = 1/ \sqrt{ 8 \pi n_0 a}$
\cite{lun97},
%\begin{equation}
%  \xi =  \frac{1}{ \sqrt{ 8 \pi \rho_0 a_s}}.
%\label{heal}
%\end{equation}
%
where the ground state density $ n_0 $ is evaluated at the vortex
position.
%When introducing dipole-dipole interaction this parameter
%becomes meaningless.
However, in the presence of dipole-dipole interactions this parameter
does no longer provide a good estimate of the vortex core size.

%***** MARTA *****
%La healing length se encuentra igualando el termino cin\'etico con
%el de interaccion en un sistema uniforme. Se podria hacer algo
%análogo con el t\'ermino dipolar?

A possible characterization has been given by O'Dell and Eberlein
\cite{dell07} assuming a variational ansatz for the vortex density
profile:
\begin{equation}
  \rho_v(r_\perp,z) =  n_0 \left(1- \frac{r_\perp^2}{R^2} - \frac{z^2}{Z^2}\right)
  \left(1- \frac{\beta^2}
{ r_\perp^2 + \beta^2}\right)\, , \label{beta}
\end{equation}
where $\beta$, $R$ and $Z$ are variational parameters that describe
the size of the core, and the radii in the transversal and axial direction, 
respectively. We want to note that for a pure $s$-wave condensate in the 
Thomas-Fermi approximation the
product of the first two factors in Eq.~(\ref{beta})
correctly describes the ground state density,
identifying $ R $ and $Z $ with the Thomas-Fermi radii. On the other
hand, the third
factor in the above formula satisfactorily models the vortex core shape.
Clearly, the quotient between the vortex
and ground state densities in the TF limit is zero at the vortex position and unity outside
the vortex core.
In particular, it is easy to verify that
 $ \beta $ corresponds to the radius
at which this quotient is equal to $ 1/2$ at the $z=0$ plane.

\begin{figure}
\epsfig{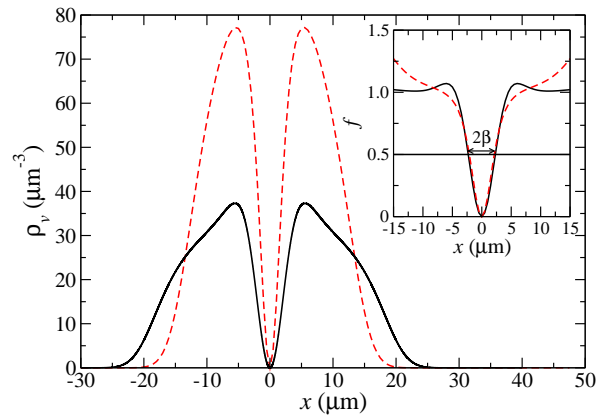}
\caption{(Color online)
Vortex density profile as a function of $x$ at $y=z=0$ for BECs with contact ($a=5 a_B$) plus dipolar interaction (solid line) and only contact interaction (dashed line). The parameters are the same as in Fig.~\ref{density0}.
Inset: $f$ as a function of the distance to the
vortex core for both cases; the value of the core radius $\beta$ is also indicated.
\label{vortex}
 }
\end{figure}

From the calculated vortex and ground state  
densities $\rho_v({\bf r})$ and $ \rho_0({\bf r})$, we propose as a definition of the core radius $\beta$ the
$r_\perp$ value in the $z=0$ plane that satisfies:
\begin{equation}
 f(r_\perp=\beta, z=0)= \frac{ \rho_v (r_\perp=\beta, z=0)}{ \rho_0(r_\perp=\beta, z=0)} =
 \frac{ 1 }{ 2 } \,. \label{EqF}
\end{equation}
This generalizes the definition given in Ref.~\cite{dell07}.
We show in Fig.~\ref{vortex} the density profile 
as a function of $x$ at $y=z=0$, $\rho_v(x,y=0,z=0)$, corresponding to a vortex
state of a condensate with dipolar plus contact interactions (solid line) and with only contact interaction (dashed line). Here 
$a = 5 a_B$ and the parameters are the same as in Fig.~\ref{density0}:
$N=10^5$, $\lambda=11$ and $\omega_\perp=8.4\times2\pi$ s$^{-1}$,
which correspond to $D=50$.
In the inset, the ratio $f$ is depicted as a function of the distance
to the vortex core for both cases. We can see that $f$ does not take the value $f=1$ outside the core. This is due to the fact that we are not in the TF regime and the structure of the BEC surface becomes important. When no dipolar effects are considered the deviation from the value $f=1$ is larger. Nevertheless, Eq.~(\ref{EqF}) still provides a good definition of the vortex core.

\begin{figure}
\epsfig{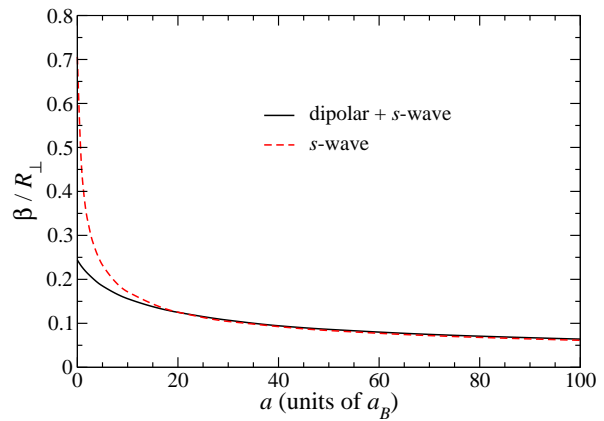}
\caption{(Color online)
Ratio of the vortex core size ($\beta$) and the radial size of the 
condensate with respect to the scattering length for a disk-shaped 
condensate with $\lambda=11$, $N= 10^5$ and $D=50$.
Solid curve: $s$-wave plus dipolar 
interactions. Dashed curve: only $s$-wave interactions.
\label{core}
 }
\end{figure}

Figure \ref{core} shows the ratio ($\beta / R_{\perp}$) of the vortex 
core size to the radial size of the disk-shaped condensate of 
Fig.~\ref{vortex} for different values of the $s$-wave scattering 
length.
The effect of the dipolar interactions for large scattering lengths
is to slightly increase the relative value of the 
core size with respect to the radial size of the condensate above the value 
of the pure $s$-wave case, as 
already discussed~\cite{dell07}. 
However, when $a<20 a_B$, 
the ratio $\beta / R_{\perp}$ is larger for a condensate with dipolar
interactions than in a pure contact interaction BEC.
This is due to the small 
repulsive interaction brought by the dipole-dipole potential, which 
becomes noticeable only for small scattering lengths. This interaction has
the effect of decreasing the central density of the condensate as compared 
to the pure $s$-wave case and this causes a smaller core radius 
and a broader ground state.

\section{ Energy barrier }\label{Barrier}

%Mir\'e (no se si es obvio) que en el caso interactuante no hay barrera.
%La altura de
%la barrera, nuevamente, estará asociada al grado de interaccion?

The nucleation of a vortex is associated with the existence of an
energy barrier in the configuration space between the initial
vortex-free state and the final vortex state. Therefore, the system
has to overcome this barrier in order to nucleate a stable vortex.
It is usually found that the vortex is nucleated at the boundary by surface
excitations and it is stable at the center of the trap for
rotational frequencies $\Omega \geq \Omega_c$. Then,
the formation energy of a vortex can be obtained by calculating the
energy of a single off-center vortex 
%in the rotating frame
as a function of the vortex core position.% \cite{Kraemer2002}.

We have generalized the ansatz of an axially symmetric
vortex line
Eq.~(\ref{wf-ansatz}) to describe an off-center vortex at
${\bf{r}}_{v}=(x_v,y_v,z)$ \cite{Jezek08,Fetter08}:
\begin{equation}
\psi({\bf{r}})= \psi_0({\bf r}) \,\frac{(x-x_v)+i \,
(y-y_v)}{\sqrt{(x-x_v)^2+(y-y_v)^2}} \,. \label{psi_v}
\end{equation}
Here the phase has been written in cartesian coordinates and
corresponds to the azimuthal angle around the shifted position of
the vortex core ${\bf{r}}_{v}$.

To calculate the energy 
%in the rotating frame
of an off-center
vortex line at a fixed distance $d=\sqrt{x_v^2+y_v^2}$ from the $z$ axis,
$E(d,\Omega)$, we have solved the GP equation in the
rotating frame taking Eq.~(\ref{psi_v}) as initial wave function. In order to obtain the solution for the displaced vortex, which is not a minimum of Eq.~(\ref{gp}), we have imposed that during the minimization process the initial nodal planes are kept constant, that is:
\begin{eqnarray}
 &&\text{Re}\left[\Psi(x_v,y,z)\right]=0 \quad \forall\, y,z \\
 &&\text{Im}\left[\Psi(x,y_v,z)\right]=0 \quad \forall\, x,z \ .
\end{eqnarray}
With this method, the quantization of the circulation is assured in all cases, but the solutions are restricted to the case of straight vortex lines. An upper bound to the formation energy of the vortex is then obtained from the difference $\Delta E(d,\Omega)=E(d,\Omega)-E_0$.

%the condition (\ref{psi_v}) during the minimization process. The difference between this energy and the energy of the ground state represents the formation energy of the vortex: $\Delta E(d,\Omega)=E(d,\Omega)-E_0$.

\begin{figure}
\epsfig{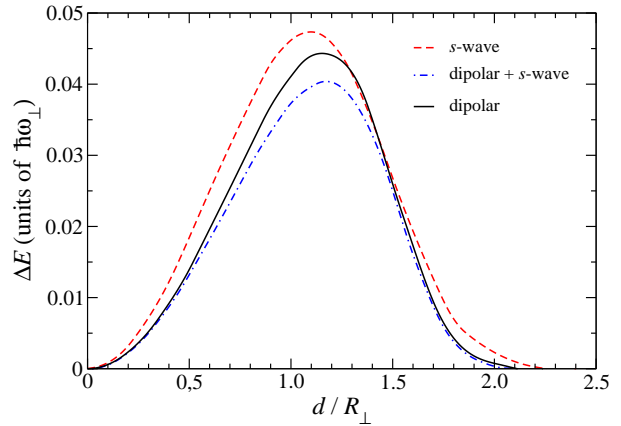}
\caption{(Color online) Energy barrier for the nucleation of a vortex in the
rotating frame at $\Omega=\Omega_c$ as a function of the vortex
displacement from the center. The dashed line corresponds to
the pure contact interaction BEC (with $a=5 a_B$), the solid
line corresponds to a pure dipolar BEC, and the dash-dotted
line to a condensate with contact plus dipolar interactions.
}
\label{barrier-critica}
\end{figure}

%
We plot in Fig.~\ref{barrier-critica} the vortex formation energy
as a function of the vortex displacement from the center, corresponding
to the same condensate as in Fig.~\ref{vortex} rotating 
at the critical rotational frequency $\Omega_c$. The distance of the vortex core to the symmetry axis is expressed in units of $R_\perp$ of the corresponding ground state.
%We assume that the vorticity line is pinned at ${\bf{r}}_{\perp,v}=(0,d)$ and express the distance of the vortex core to the symmetry axis in units of $R_\perp$ of the ground state.
%the rms-radius in the transverse plane of the vortex free
%condensate.%:
%
%$R_\perp=(\int r^2 |\psi_0({\bf r})|^2 d{\bf r})^{1/2}$.
%

The dashed line corresponds to the pure contact interaction
BEC (with $a=5 \,a_B$), the dash-dotted line to a condensate with
contact plus dipolar interactions, and the solid
line corresponds to a pure dipolar BEC ($D=50$). Each curve has been
calculated at the corresponding critical angular velocity
(see Table I).
%that is at $\Omega_c=0.4486\omega_\perp, 0.2489\omega_\perp$ 
%and $0.2491\omega_\perp$,
%for the pure contact interaction, both contact and dipolar
%interactions, and pure dipolar condensate, respectively. 
It is interesting to note that even though $R_{\perp}$ is different for each curve, the three critical barriers have the same
qualitative behavior as a function of the dimensionless 
displacement of the vortex core $d/R_\perp$.
%are very similar when they are expressed as a
%function of the dimensionless displacement of the vortex core
%$r_v/R_\perp$. 
At $\Omega_c$ the centered vortex state and the
vortex-free state have the same energy but they are
separated by an energy barrier. The maximum of the barrier height $\Delta E$
is located around $d_{max}/R_\perp \sim 1.1$ for the contact interaction BEC and around $d_{max}/R_\perp \sim 1.2$ for the other cases. Since the radius of the
condensate is larger for a system with dipolar plus contact
interactions this means that, for the vortex state configuration
at the barrier maximum,
the distance between the vortex core and the $z$ axis
is larger than in the case of a pure dipolar
condensate, which turns out to be also larger than the distance in a
pure contact interaction BEC (see Table I).

For the sake of comparison we also give in Table I the barrier parameters at the critical frequency calculated in the TF approximation for a pure $s$-wave BEC. The barrier height and the position of the maximum can be obtained, respectively, from the expressions \cite{Kraemer2002}
\begin{equation}\Delta E(d_{max},\Omega)= \frac{2}{5}\Omega_c\left(\frac{3}{5}\frac{\Omega_c}{\Omega}\right)^{3/2}\end{equation}
and
\begin{equation}d_{max}=\sqrt{3}R_\perp\sqrt{1-\frac{3}{5}\frac{\Omega_c}{\Omega}}\ ,\end{equation}
where we have used that the TF radius in the radial direction and the corresponding root-mean-square value are related by $R_{TF}=\sqrt{3}R_\perp$. The critical frequency can be evaluated using \cite{lun97}
\begin{equation}\Omega_c = \frac{5\hbar}{6mR_{\perp}^2} \ln\left(\frac{0.671\sqrt{3}R_{\perp}}{\xi}\right)\ .\end{equation}
Although the case we have considered does not really correspond to the TF limit ($Na/a_{ho}\approx8$), we can see in Table I that this approximation can be used to obtain an estimate of the energy barrier, especially the location of its maximum, which matches very well the numerical result.

%When the vorticity is pinned at large distances $r_v/R_\perp > 1$ the
%vortex line is located at a small density region and therefore the
%numerical calculation of the circulation is not as accurate
%as in the internal regions of the condensate. For this reason the
%right side of the energy barriers are shown as dotted lines 
%just to guide the eye.

\begin{table}
\caption{\label{tab:1} Characteristics of the energy barriers shown in Fig.~\ref{barrier-critica}.
}
\begin{ruledtabular}
\begin{tabular}{l|ccccccccc}
Interaction & $R_\perp (\mu m)$ & $\Omega_c (\omega_\perp)$ &
$d_{max} (\mu m)$ & ${d_{max}}/{R_\perp}$ & $\Delta E (\hbar \omega_\perp)$ \\
%$r_v^{max} (\mu m)$ & $\frac{r_v^{max}}{R_\perp}$ & $\Delta E (\hbar \omega_\perp)$ \\
\hline
$a=0$ & $11.62$ & $0.25$ & $13.48$ & $1.16$ &
$0.044$ \\
$\mu= 6 \mu_B$ & & & & & \\
\hline
$a=5 a_B$ & $12.44$ & $0.25$ & $14.55$ & $1.17$ & $0.040$ \\
$\mu= 6 \mu_B$ & & & & & \\
\hline
$a=5 a_B$ & $9.16$ & $0.45$ & $9.98$ & $1.09$ & $0.047$ \\
$\mu=0$ & & & & &\\
\hline
$a=5 a_B$ & $10.84$ & $0.38$ & $11.87$ & $1.09$ & $0.071$\\
(TF) & & & & &
%
%$N_2$ & $  6$& $6.81$& $7.69$& $ 8.63$& $9.80 $  & & & &
\end{tabular}
\end{ruledtabular}
\end{table}

\begin{figure}
\vspace*{0.5cm}
\epsfig{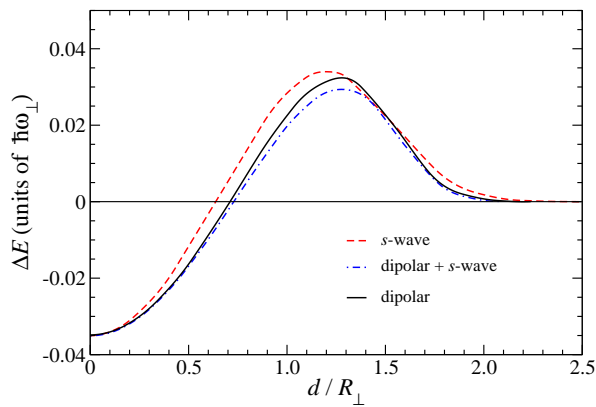}
\caption{(Color online) Vortex formation energy as a function of the vortex
displacement from the center at angular velocities $\Omega >
\Omega_c$. The dashed line corresponds to
the pure contact interaction BEC (with $a=5 a_B$), the solid
line corresponds to a pure dipolar BEC, and the dash-dotted
line to a condensate with contact plus dipolar interactions.
 }
 \label{barrier}
\end{figure}

We plot in Fig.~\ref{barrier} the same curves as in
Fig.~\ref{barrier-critica} but calculated at larger angular velocities $\Omega > \Omega_c$, namely $\Omega= 0.48~ \omega_\perp$
%$\Omega= 0.4836\omega_\perp$ 
for the pure contact interaction BEC with $a=5 \,a_B$ (dashed line)
%$\Omega= 0.2839\omega_\perp$
%$\Omega= 0.28 \omega_\perp$ for the condensate with both contact and
%dipolar interactions (dashed line), and $\Omega= 0.2839\omega_\perp$ 
%for the pure dipolar BEC (dash-dotted line).
and $\Omega= 0.28 ~\omega_\perp$ for both the pure dipolar BEC 
(solid line) and the condensate with the two types of interactions acting simultaneously (dash-dotted line).

As expected, for rotational frequencies larger than the
corresponding critical one, the state with a centered vortex is
preferable. However, the nucleation of the
vortex is inhibited by the barrier separating the vortex-free state
from the energetically favored vortex state
which corresponds to the minimum energy configuration, at
$d/R_\perp=0$. %One can see from Figs.~\ref{barrier-critica} and
%\ref{barrier} that when the angular velocity
%of the trap is increased, the position of the barrier height is
%shifted towards the condensate surface and the energy height
%decreases.

Finally, we have numerically checked that when we start the minimization procedure with an off-center, unpinned vortex line
located at distances larger than
the position of the barrier maximum $d >
d_{max}$, the system converges to the vortex-free state. On the contrary, an initially
 off-center vortex located
at $d < d_{max}$ converges to a centered vortex state.

%
%\begin{table}
%\caption{\label{tab:1} Equilibrium radii of the rings of vortices
%obtained for the two number of particles we have
%considered: $ N_1 = 2.5 \times 10^{5}$ and $ N_2 =  10^{6}$.}
%\begin{ruledtabular}
%\begin{tabular}{l|ccccccccc}
%$N_v$ & 1 &2 & 3 & 4 & 5 & 6 & 7 & 8 & 9 \\
%\hline
%$N_1$ &  $  6$& $6.05$& $6.18$& $6.41$& $6.67$& $7.05$& $7.33$& $7.86$& $8.72$\\
%$N_2$ & $  6$& $6.81$& $7.69$& $ 8.63$& $9.80 $  & & & &
%\end{tabular}
%\end{ruledtabular}
%\end{table}

\section{Summary and concluding remarks }\label{Conclusions}

In this paper we have adressed singly quantized vortex states
in dipolar pancake shaped condensates.
%We have fixed the dipole-dipole interaction and have changed the $s$-wave scattering length to explore different regimes experimentally available. For small values of $a$ the dipolar interaction dominates, becoming a pure dipolar condensate when $a=0$. On the contrary, when $a$ is large the dipolar interaction is negligible \cite{dell07}.
By fixing the dipole-dipole interaction and scanning the $s$-wave scattering length from $a= 100\,a_B$ to zero, we have been able to explore the vortex states in a dipolar condensate in three different regimes: when the two-body interaction is governed by the contact potential, when it is the dipole-dipole interaction that controls the properties of the gas and, finally, the region where both interactions are comparable.

%The ground state of the dipolar condensate has been revisited: we have discussed its structure and have shown that the Gross-Pitaevskii theory predicts new configurations which are not found in $s$-wave condensates and which are stable. 
We have reviewed some properties of the ground state of the dipolar condensate previously obtained in the bibliography. In particular, we have discussed its structure and have numerically solved the GP equation to obtain the new stable configurations which are not found in $s$-wave condensates.
We have also discussed that the effect of the dipolar interaction on the effective size of the condensate depends strongly on the geometry of the confining potential. In a spherical trap, this anisotropic interaction tends to increase the size of the condensate in the direction perpendicular to the magnetization axis while reducing it in the parallel direction. However, we have found that in a pancake trap the radius is increased in both directions in such a way that the cloud aspect ratio also increases.

We have calculated the singly quantized vortex states of dipolar condensates in two different pancake configurations. We have obtained an excellent agreement with the results in Ref.~\cite{dell07} and have extended the study for scattering lengths as small as possible. We have seen that the effect of the dipolar interactions is to reduce the critical frequency for the nucleation of a vortex as compared to the $s$-wave case, and that a maximum in $\Omega_c$ appears at low scattering lengths (becoming more important near collapse). We have also obtained the structure of the vortex core and have shown that there exists a certain value of scattering length below which the dipolar effects become dominant.

Finally, we have characterized the energy barrier which has to be overcome to nucleate a vortex, both at the critical frequency and above it. To the best of our knowledge, this is the first study where energy barriers have been addressed in dipolar condensates. We have compared three different cases: a pure $s$-wave condensate, one with only dipolar interactions  and a BEC with both types of interactions. The barriers as a function of the dimensionless vortex displacement exhibit a very similar shape with the following slight differences: for condensates with dipolar interactions, they are narrower and lower, indicating that it is energetically less expensive to nucleate a vortex in a dipolar BEC than in one with only contact interactions.

\acknowledgments We thank Manuel Barranco for helpful discussions. This work has been performed under Grant No.
PIP~5409 from CONICET and FIS2008-00421 from MEC (Spain). % and
%2005SGR-00343 from Generalitat de Catalunya.
M. A. is supported by the Comission for Universities and Research of the Department of Innovation, Universities and Enterprises of the Catalan Government and the European Social Fund.

\appendix
\section{}\label{Virial}
In this appendix we obtain the virial theorem for a dipolar BEC from the density functional Eq.~(\ref{ed}), by using the principle of scale invariance. The virial theorem results from the homogeneity properties of the kinetic and potential components of the energy of the many-body system with respect to a scaling transformation that preserves the normalization.

%It is well known that the virial theorem can be obtained from scaling laws applied to the different terms of the energy functional of a system. Our purpose in this appendix is to find the expression of the virial theorem for a condensate of dipoles. 

Considering the following transformation $\mathbf{r}\rightarrow\nu\mathbf{r}$, the condensate wave function scales as $\Psi(\mathbf{r})\rightarrow\Psi_\nu(\textbf{r})=C\Psi(\nu\mathbf{r})$, where $C$ is a normalization constant. The principle of scale invariance ensures that the norm of the wave function is preserved, that is
\begin{equation}\int d\mathbf{r}\left|\Psi({\bf r})\right|^2=
\int d\mathbf{r}\left|\Psi_\nu({\bf r})\right|^2= 
|C|^2\int d\mathbf{r}\left|\Psi(\nu{\bf r})\right|^2=N \ ,\end{equation} %2     \nu^{3/2}\psi({\bf r})\ .\]
which gives $C=\nu^{3/2}$. Using this result, it has already been shown (see, for instance, Ref.~\cite{Jezek2004}) that the kinetic, harmonic potential and contact interaction terms in the energy functional (\ref{ed}) scale as
%It has been shown (see for instance \cite{Jezek2004}) that the three first contributions to the energy functional (\ref{ed}) scale following the rules: 
\begin{eqnarray}
&& E_{\text{kin},\nu}  = \nu^2 E_{\text{kin}}\ ,\\
&& E_{\text{trap},\nu}  = \frac{1}{\nu^2}E_{\text{trap}} \ , \\
&& E_{\text{int},\nu} = \nu^3E_{\text{int}}\ . 
\end{eqnarray}

For the contribution of the dipolar energy to the functional, one can proceed in the same way and obtain its scaling law
\begin{eqnarray}E_{\text{dip},\nu} &&\!\!\!= \frac{1}{2}\frac{\mu_0\mu^2}{4\pi}\!\!\int\! d{\bf r} d{\bf r'} \frac{|\psi_\nu({\bf r})|^2|\psi_\nu({\bf r'})|^2}{|{\bf r}-{\bf r'}|^3}(1-3\cos^2\theta)\nonumber\\    &&\!\!\!\!\!\!\!\!\!\!\!\!\!\!\!\!\!= \frac{1}{2}\frac{\mu_0\mu^2}{4\pi}\!\!\int\! d{\bf r} d{\bf r'} \frac{\nu^6|\psi(\nu{\bf r})|^2|\psi(\nu{\bf r'})|^2}{|{\bf r}-{\bf r'}|^3}(1-3\cos^2\theta)\nonumber\\
\!\!\!\!=\!\frac{1}{2}&&\!\!\!\!\!\!\!\frac{\mu_0\mu^2}{4\pi}\nu^3\!\!\int\! d(\nu{\bf r}) d(\nu{\bf r'}) \frac{|\psi(\nu{\bf r})|^2|\psi(\nu{\bf r'})|^2}{|\nu{\bf r}-\nu{\bf r'}|^3}(1-3\cos^2\theta)\nonumber\\
&&\!\!\!\!\!\!\!\!\!\!\!\!\!\!\!\!\!=\nu^3 E_{\text{dip}}\ .
\end{eqnarray}
Therefore the total energy of the system can be written as
\begin{equation}E_\nu= \nu^2 E_{\text{kin}} + \frac{1}{\nu^2}E_{\text{trap}} + \nu^3E_{\text{int}} + \nu^3 E_{\text{dip}}\ .\end{equation}
Imposing the equilibrium condition
\begin{equation}
 \left.\frac{dE_v}{d\nu}\right|_{\nu=1} = 0
\end{equation}
%that the energy must be a minimum and that in the equilibrium position $\nu=1$, %
%the stationarity condition of the energy against dilation:}
%\begin{equation}
% 0=\left|\frac{\partial E_\nu}{\partial \nu}\right|_{\nu=1}\ ,
%\end{equation}
one finds the virial theorem for a dipolar condensate, Eq.~(\ref{EqVirial}).
%\begin{equation}
%\begin{equation} 2E_{\text{kin}} - 2E_{\text{trap}} + 3E_{\text{int}} + 3E_{\text{dip}} = 0 \ .\end{equation}
%\end{equation}
Note that the energy term $E_L$ in Eq.~(\ref{ed}) does not depend on the scaling parameter, and thus it does not contribute to the virial expression.


\begin{thebibliography}{99}
\bibitem{foot1} Even though 3D effects on vortex states should be more evident in cigar shaped configurations than in pancake ones, the stability of the condensate with $a\rightarrow0$ is far more assured in the latter. Therefore, in order to reach stable configurations in the limit $a=0$ we restrict ourselves to pancake geometries.
\bibitem{foot2} If another definition for the radius is used (for instance, the FWHM) the actual values of $R_\perp$, $R_z$ and $\kappa$ are different but we checked that they predict the same behavior for the cloud aspect ratio.
\end{thebibliography}
\end{document}